\documentclass{Interspeech2024}
\usepackage{multirow}
\usepackage{booktabs} 
\newcommand{\LaRedNewName}{REWIND\xspace}

\interspeechcameraready

\title{How Private is Low-Frequency Speech Audio in the Wild?
An Analysis of Verbal Intelligibility by Humans and Machines}

\name[affiliation={1,2}]{Ailin}{Liu}
\name[affiliation={2}]{Pepijn}{Vunderink}
\name[affiliation={2}]{Jose Vargas}{Quiros}
\name[affiliation={2}]{Chirag}{Raman}
\name[affiliation={2}]{Hayley}{Hung}

\address{
  $^1$RWTH Aachen University, Germany\\
  $^2$Delft University of Technology, Netherlands}
\email{ailin.liu@rwth-aachen.de, p.j.vunderink@student.tudelft.nl, j.d.vargasquiros@tudelft.nl, c.a.raman@tudelft.nl, h.hung@tudelft.nl}


\begin{document}

\maketitle

\begin{abstract}
   Low-frequency audio has been proposed as a promising privacy-preserving modality to study social dynamics in real-world settings. To this end, researchers have developed wearable devices that can record audio at frequencies as low as 1250 Hz to mitigate the automatic extraction of the verbal content of speech that may contain private details.
   This paper investigates the validity of this hypothesis, examining the degree to which low-frequency speech ensures verbal privacy. It includes simulating a potential privacy attack in various noise environments. Further, it explores the trade-off between the performance of voice activity detection, which is fundamental for understanding social behavior, and privacy-preservation. The evaluation incorporates subjective human intelligibility and automatic speech recognition performance, comprehensively analyzing the delicate balance between effective social behavior analysis and preserving verbal privacy. 
\end{abstract}

\section{Introduction}
Speech, as a fundamental modality, serves as a rich source for studying various paralinguistic aspects of human behavior, encompassing elements such as prosody, intonation, and rhythm \cite{VINCIARELLI20091743, myers2019pushing, guyer2019speech}. Analyzing these features not only provides insights into emotional states, social dynamics, and communication patterns \cite{5549893, hung2010estimating} but also contributes to advancements in fields such as linguistics, psychology, and human-computer interaction. However, analyzing human behavior through speech analysis presents a significant challenge in ensuring privacy, especially in real-world settings where individuals may inadvertently disclose sensitive information in natural conversations. 

Striking a balance between extracting valuable paralinguistic insights and preserving privacy becomes paramount in ethical and responsible research practices, especially in real-life applications. One promising strategy is to use low-frequency audio recordings in smart badges \cite{raman2022conflab, 8259434} which allows for analysis of paralinguistic features while mitigating the risk of inferring verbal content. Recording at a low frequency makes it possible to infer essential nonverbal elements of social and emotional behavior such as turn-taking and prosodic features without compromising the privacy of the verbal content. This is particularly relevant in the wild, where unscripted and spontaneous interactions occur, reflecting genuine human behavior.

This study investigates the feasibility of leveraging low-frequency audio recordings captured in real-world settings to infer plausible verbal content, which could be any words perceived artificially or by human listeners. Our primary emphasis is on verbal privacy, as nonverbal features, including gender and personal attributes \cite{wolfson1989social}, are commonly investigated in the context of social dynamics.
Employing automatic techniques, including speech-to-text conversion and bandwidth-extension methods, we aim to explore the potential of extracting meaningful insights from these recordings while safeguarding the privacy of individuals involved in the conversations.
While other technical strategies exist for preserving semantic privacy post-recording, low-frequency audio promises to provide users with agency in providing informed consent. Informing users beforehand that their audio is to be recorded at low frequencies for privacy sensitivity reasons is crucial for empowering users to choose how their data is used for research and promoting social and emotional well-being in this research space.
\begin{figure}[t]
\centering
\includegraphics[width=0.48\textwidth]{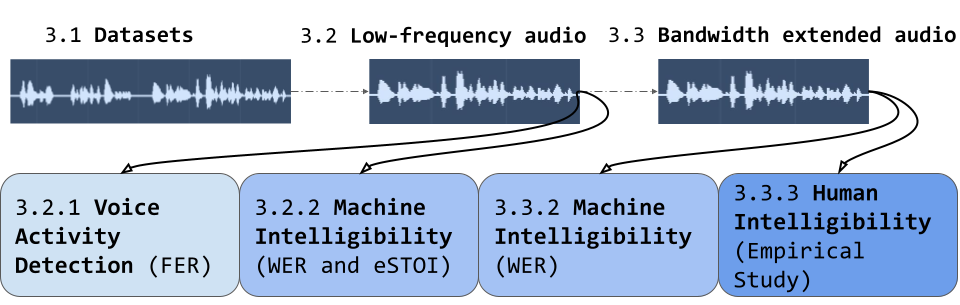}
\caption{Overview of the study. From datasets with and without mingle setting (Section 3.1), we process the audio samples into low-frequency speech audio (Section 3.2) and bandwidth-extended low-frequency speech audio (Section 3.3).}
\label{fig:mathod}
\end{figure}
\section{Related work}
An approach for analyzing turn-taking using voice activity detection (VAD) in privacy-sensitive speech is to extract audio features \cite{articlerw} which cannot be used to reconstruct intelligible verbal speech content. Applying the Principal Component Analysis method to an audio spectrogram has been proposed to detect non-speech activity and prevent speech reconstruction \cite{10.1145/2030112.2030163}. Moreover, encryption methods on privacy-sensitive audio are available to hide verbal content \cite{inproceedingsrw} for speaker segmentation tasks or obfuscation in urban sound recording \cite{cohen2019voice}. Low-frequency audio \cite{8594870} has been used for group gender classification under privacy-sensitive speech. Sound shredding \cite{10.1145/2699343.2699366}, slicing \cite{maouche2022enhancing}, subsampling \cite{articless}, and degradation \cite{10.1145/3379503.3403551} are methods to mutate the raw sound which makes it difficult to recover the verbal content of the original recording but maintain some acoustic features of it. Alternatively, replacing the original data with artificial speech generated from Generative Adversarial Network (GAN) architectures \cite{8771098} is used. Also, some work proposes using speech embeddings to preserve privacy \cite{teixeira2022towards}. The advantage of utilizing low-frequency audio lies in its transparent nature. Users can conceptualize the sound of low-frequency recordings or actively listen to them, gaining a clear understanding that their privacy is safeguarded. In contrast to alternative methods like encryption, where users must rely on trust in researchers to ensure data usage aligns with consent, low-frequency recordings eliminate certain potential misuses, because specific information is inherently absent from the recorded signal, providing users with a tangible and reassuring layer of privacy.
 
\section{Analysis of Low-Frequency Audio}
We examined the performance of low-frequency audio on VAD, automatic speech recognition (ASR), and extended short-term objective intelligibility (eSTOI) \cite{7539284}. To apply an intuitive attack, we used bandwidth-extension (BWE) methods to potentially improve the intelligibility. Bandwidth extension of audio is a task aiming to enhance speech quality over narrow-band telephone connections by extrapolating higher frequencies missing in the low-resolution input. To assess the effectiveness of the potential attack, the human and machine intelligibility of the bandwidth-extended audio is measured.
Figure \ref{fig:mathod} shows an overview of our study. In section 3.1 we present three audio datasets that were used, each being recorded in different noise settings. In section 3.2 we make a comparison across sample rates of VAD performance (3.2.1) and automatic speech intelligibility. In section 3.3 we extend the analysis to bandwidth-extended audio, both for machine speech intelligibility (3.3.2) and speech intelligibility by humans (3.3.3).

\subsection{Datasets}

We used three datasets in our study: Pop-glass \cite{popglass, popglass2}, VCTK \cite{https://doi.org/10.7488/ds/2645}, and \LaRedNewName \cite{quiros2024rewind}. Pop-glass and \LaRedNewName were recorded in mingling environments. In Pop-glass, speech is mainly in English, while in \LaRedNewName speech is mainly in Dutch. VCTK, on the other hand, was recorded in clean audio conditions and is in English. Further details about each dataset are described below.

\noindent\textbf{Pop-glass} consists of $32$ people who participated in a mingling event with the official spoken language in English. Each recording was approximately 1 hour long. Every participant wore an omnidirectional Lavalier microphone attached to the face. The original frequency of the audio is 44.1 kHz. $27$ out of $32$ recordings are included in our study after filtering out completely silent audio and audio from malfunctioning microphones.

\noindent\textbf{VCTK} is an English multi-speaker corpus provided in the CSTR voice cloning toolkit. Each speaker reads a different set of sentences from a newspaper article in a quiet and single-speaker setting. The original frequency of the audio is 48 kHz. The audio of a female speaker is used which aligns with the open-sourced pre-trained model available for speech enhancement on VCTK.

\noindent\textbf{\LaRedNewName} contains personal audio recorded from individual microphones. The setting is a professional networking event with around 100 attendees. 43 consented to wear a microphone and from this data, 16 people's audio data were selected for our experiment to ensure a diverse set of speakers. The microphone used and the original recording frequency is the same as Pop-glass. Most of the audio is in Dutch, although sometimes English is also spoken.

\subsection{An analysis of low-frequency speech audio}

During the analysis, the main motivation was to understand how the frequency of input speech affects the state-of-the-art VAD and open-sourced ASR systems.

\subsubsection{Voice activity detection}

In this study,  we used rVAD \cite{TAN20201} for the VAD task. It is an unsupervised segment-based method and is compared favorably with some existing methods, such as Kaldi Energy VAD \cite{article} and VQVAD \cite{6639066}. To evaluate the performance of rVAD on different sample rates, false error rates (FER) are calculated as the ratio between the number of wrongly categorized events and the total number of actual events.
The same dataset was down-sampled to different frequencies before being used for evaluation on rVAD to test performance across frequencies. 27 audio samples (all samples are from different participants) from Pop-glass and 6 audio samples from p225 of VCTK are taken into account. Pop-glass samples are cut into segments of 20 to 30 seconds from 1 hour of the mingling event. All samples were down-sampled to 300, 800, 1250, 2000, 3200, 5000, 8000, and 20000 Hz, chosen to be logarithmically increasing. An order 8 Chebyshev
type I Low-pass filtering was applied before down-sampling to avoid aliasing.

\begin{figure}[t]
\centering
\includegraphics[width=0.32\textwidth]{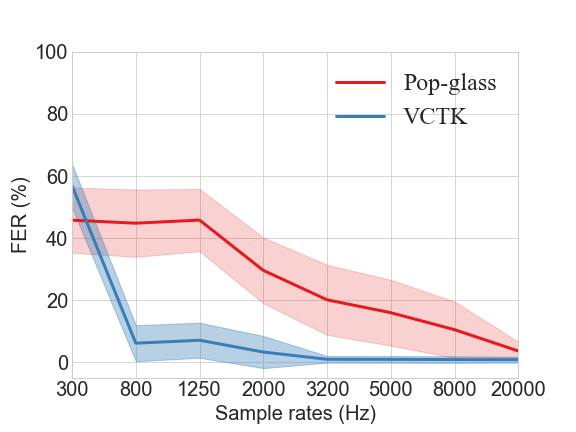}
\caption{Performances (means and standard deviations) of rVAD on different sample rates comparing to original ones}
\label{fig:vad}
\end{figure}

Figure \ref{fig:vad} shows that the FER drops dramatically when going from a 300 to an 800 Hz sample rate on the VCTK audio. A similar, though less dramatic drop in FER, is observed from 2000 Hz onwards on samples from Pop-glass. Even though the performance of VAD is sensitive to the sample rates, it is reasonable to use the VAD above 800 Hz for clean speech audio and 2000 Hz for speech audio in a mingling environment.

\subsubsection{Speech intelligibility}
In this study, automatic speech intelligibility is evaluated in terms of the performances of ASR and eSTOI. The performance of ASR evaluates whether machines can transcribe audio into text. eSTOI is an automated intelligibility listening test that compares noisy audio sources to a clean reference. 

We employed the open-sourced ASR model Whisper \cite{whisper} trained on multilingual and multitask supervised data from the web to evaluate samples in different frequencies. To evaluate the performance of the ASR model on different frequencies, word error rate (WER) \cite{wer} was calculated. Outputs of the ASR model were pre-processed by lower-case transformation, white space removal, and bag-of-words reduction before computing the WER metrics.

\begin{figure}[b]
\centering
\includegraphics[width=0.5\textwidth]{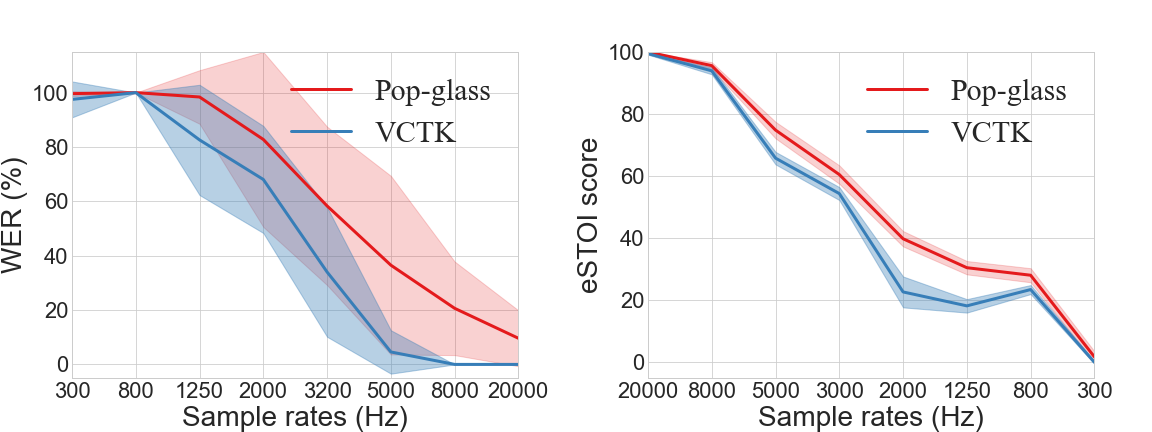}
\caption{Performances of Whisper on different frequencies compared to the ground truth transcripts and of speech intelligibility prediction from eSTOI on different frequencies compared to the original speech signals respectively.}
\label{fig:comb-asr}
\end{figure}

Figure \ref{fig:comb-asr} shows that the WER is ${\sim}10$\% for Pop-glass and is $0$\% for the samples from VCTK at $20000$ Hz. It shows that open-sourced ASR works well for high-frequency speech audio. However, the WER is higher than 97.5\% for 300 - 800 Hz VCTK audio and higher than 98\% for 300 - 1250 Hz Pop-glass audio. This indicates that ASR performance is significantly worse on low-frequency speech audio compared to high-frequency speech audio. 

 A higher score in eSTOI represents a prediction that the speech intelligibility performance will be better, compared to a given reference speech signal. The scores range between 0 and 100 (as clear as the original audio). As the eSTOI result shows, samples from both VCTK and Pop-glass maintain 20 and 40 respectively, when the sample rates are lower than 2000 Hz, compared to their high-frequency speech audio. Furthermore, there is little improvement in the intelligibility prediction between 800 and 2000 Hz. Generally, as expected, automatic speech intelligibility decreases with lower speech frequency.

\subsection{Analysis of bandwidth-extended low-frequency speech}

To understand the effect of the potential attack on low-frequency audio, we performed an analysis of ASR performance and a user study after a bandwidth-extension process on the same audio samples mentioned earlier, evaluating intelligibility by both machines and humans. 

\subsubsection{Simulating an attack via Bandwidth Extension}
By "hallucinating" higher frequencies which are absent in the low-resolution input, bandwidth-extension of audio aims to improve audio quality and intelligibility of speech. In this study, we used neural bandwidth-extension models \cite{https://doi.org/10.48550/arxiv.1708.00853}. The two models were trained on \LaRedNewName and VCTK respectively to simulate a privacy violation situation. The VCTK model trained and tested on audio from the same speaker and noise conditions simulates the lower bound of such an attack. The \LaRedNewName model simulates a more aggressive, possibly more realistic informed attack where only the noise conditions of the sample are known beforehand and exploited as part of a pre-trained BWE approach using other data (in our case, Pop-glass). The open-source VCTK model is trained on 16 kHz audio from the single speaker of VCTK and the \LaRedNewName model is trained on 8 and 5 kHz audio from multiple speakers of \LaRedNewName. Signal-to-Noise Ratio (SNR) describes the ratio of signal power to noise in the signal in the time domain, measuring BWE performance.
It represents the intensity of error in predicted audio signals to the intensity of their corresponding reference signals. As Table \ref{table:snr} shows, the higher the SNR, the better the model's performance. We selected $6$ out of $27$ samples in Pop-glass containing the minimum, 25th percentile, two medians, 75th percentile, and maximum fundamental frequency F0, as representative samples. To align with formants F1 (500 Hz) and F2 (1500 Hz) \cite{nawka1997speaker}, sample rates at 800, 1250, and 2000 Hz are evaluated in the intelligibility analysis, because F2 has been indicated for contributing the most to intelligibility \cite{han2017relative}.

\begin{table}[!htbp]
\caption{SNR result of applying models trained on VCTK and \LaRedNewName and tested on VCTK and Pop-glass audio respectively.}
\begin{center}
\begin{tabular}{ ccc} 
\toprule
 &VCTK model&\LaRedNewName model\\
\midrule
sample rates&SNR&SNR\\
\midrule
800 Hz& 2.3135&0.7124\\
1250 Hz& 3.6029&1.0873\\ 
2000 Hz& 6.5619&1.9398\\
\bottomrule
\end{tabular}
\end{center}
\label{table:snr}
\end{table}

\subsubsection{Machine intelligibility}

To evaluate how bandwidth-extended audio improves WER compared with the original low-frequency audio samples, the same model of Whisper is applied to both. 

\begin{figure}[t]
\centering
\includegraphics[width=0.5\textwidth]{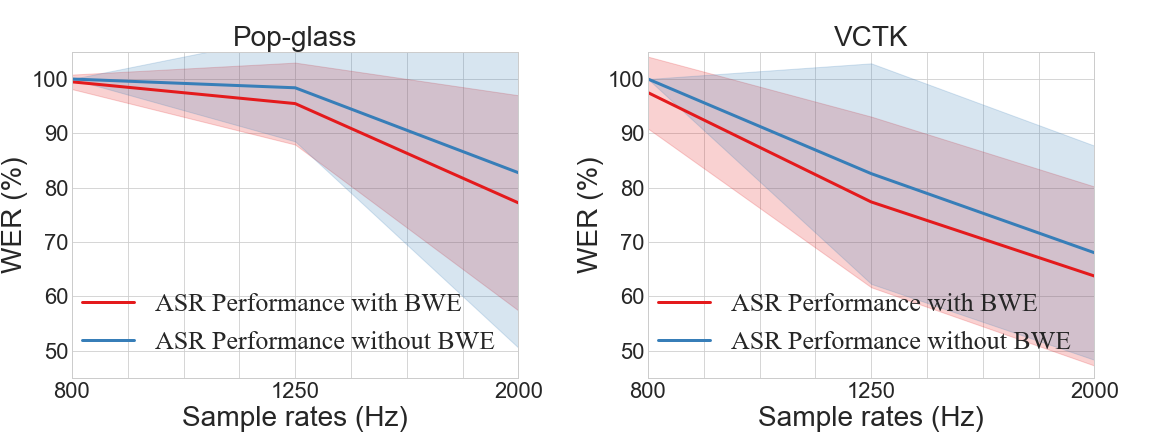}
\caption{Performances of ASR with BWE and without BWE on Pop-glass and VCTK audio respectively with sample rates, 800, 1250, and 2000 Hz compared to the ground truth transcripts}
\label{fig:bwe_asr}
\end{figure}

Figure \ref{fig:bwe_asr} shows that there is a reasonable improvement of WER achieved by the bandwidth extension models in the Pop-glass audio samples with a sample rate of 1250 or 2000 Hz and the VCTK audio samples with a sample rate of 800, 1250, or 2000 Hz. The decrease in WER can be interpreted as an improvement in automatic speech intelligibility. However, most of the words recovered from the bandwidth extension models are stop-words which might be less informative on privacy.

\begin{figure}[b]
\centering
\includegraphics[width=0.5\textwidth]{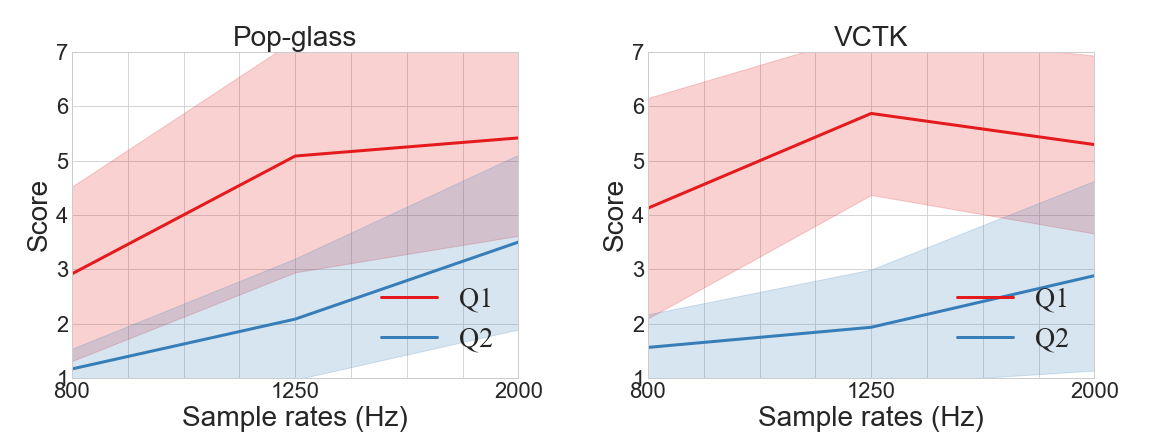}
\caption{Mean and standard deviation of Q1 and Q2}
\label{fig:q1}
\end{figure}

\subsubsection{Human intelligibility}
We conducted a perceptual experiment on speech intelligibility to investigate how much speech intelligibility is preserved in low-frequency audio. Typically, speech intelligibility is measured via rating scales \cite{rate} and word recognition tests \cite{YORKSTON1978499}. We recruited 6 participants including 4 males and 2 females. All the participants confirmed they didn't have any hearing impairment and carried out the intelligibility test inside a sound-isolated listening booth. They were asked to wear headphones for the study, but the volume was not restricted. They were permitted to increase or decrease the volume and listen to the audio samples multiple times. 14 audio samples were used; 6 of them from Pop-glass, and 8 from VCTK. After listening to each sample, they were asked to fill out a questionnaire on the intelligibility of the audio content based on a 7-point Likert scale.\\
\textit{Q1: Are you able to hear anything in the audio file?}\\
\textit{Q2: Are you able to hear speech in the audio file?}\\
\textit{Q3: Please transcribe the audio file word by word (mark all perceived but not recognized words with a character X).}\\
\textit{Q4: Do you hear more than one speaker in the conversation? If you can, state roughly how many speakers in the conversation you think there are.}

\begin{figure}[b]
\centering
\includegraphics[width=0.5\textwidth]{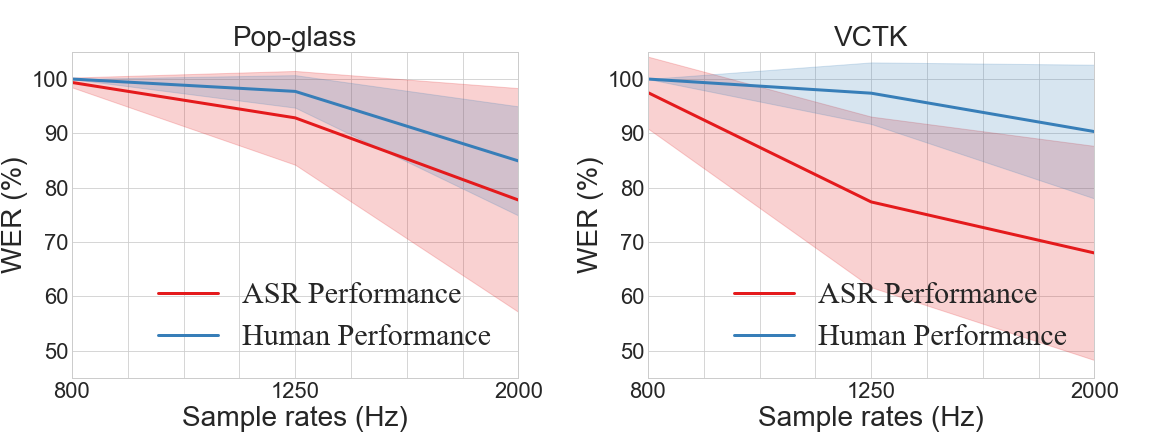}
\caption{WER Performances of ASR and Human perception on Pop-glass and VCTK bandwidth-extended audio respectively with sample rates, 800, 1250, and 2000 Hz compared to the ground truth transcripts}
\label{fig:asr_human}
\end{figure}

Q1 and Q2 are measured on a Likert scale (1 to 7, 1 being “Not at all” and 7 being “Very clearly”). 
\autoref{fig:q1} illustrates the results. For both datasets, higher sample rates correlated with higher speech intelligibility scores. However, 2000 Hz audio is not perceived as significantly clearer than 1250 Hz audio. Q3 is measured by WER first and Figure \ref{fig:asr_human} shows when transcribing low-frequency speech recordings, humans perform marginally worse than the open-sourced ASR. Q3 is also analyzed by other metrics in the next paragraph. Q4 is posed for gaining contextual information about whether the main speaker was transcribed. 
The number of recognized speakers in Pop-glass followed a mean and std of $0.67 (0.62)$ at 800 Hz; $1.417 (0.64)$ at 1250 Hz; and $1.58 (0.64)$ at 2000 Hz. For VCTK, at all sample rates, the means and std were found to be $1 (0)$. 
Consequently, the results on VCTK are representative of the primary speaker. The results of Pop-glass indicate that cross-talk constitutes another source of privacy threat; the question of whose information is leaked is beyond the scope of the present analysis focusing on the verbal intelligibility of low-frequency audio but warrants future investigation.\\

\noindent\textit{Metrics for Human Speech Intelligibility:} 
Beyond evaluating the WER of the transcripts, as shown in Table \ref{table:metr}, we introduce the following metrics to measure human speech intelligibility: 

 \noindent\textbf{The number of recognizable words} represent words that participants can write down. The number of recognizable words evaluates how many words could be perceived regardless of whether they were truly spoken. 

\noindent\textbf{The number of perceivable words}
 estimates how many words are perceived, including recognizable words and those that cannot be spelled by participants, but the beginning and the end of which can be identified. It provides a good insight into whether the audio can be used to detect multiple potential words.

\noindent\textbf{The ratio of recognizable and perceivable words} measures how many words are recognized from all words that were perceived to have been spoken in the audio sample.

\noindent\textbf{The longest chain of recognizable words}
of the audio samples has been chosen to determine whether the recognized words are located randomly or continuously in a sentence. Continuous recognizable words tend to provide more information than corpora located randomly in sentences. 

\noindent\textbf{Pairwise cosine similarity} 
of each audio sample was chosen to measure the similarity between transcripts of two participants listening to the same audio. A higher pairwise cosine similarity means more identical words are shared in the transcripts of participants. Audio at $2000$Hz in a mingle setting has a significantly higher pairwise cosine similarity than others. It reveals many words that can be identified at 2000 Hz but not at 1250 Hz for all participants. Thus, 1250 Hz is a reasonable threshold that blocks most of the intelligible verbal content during mingling. 
\begin{table}[t]
\caption{Result of the proposed metrics for evaluating speech intelligibility by humans.}
\centering
 \resizebox*{\linewidth}{!}{\begin{tabular}{ cc|cc|cc}
 & &\multicolumn{2}{|c|}{\# recognizable}&\multicolumn{2}{|c}{\# perceivable}\\
\hline
dataset&frequency&Mean&SD&Mean&SD\\
\hline
\multirow{3}{*}{Pop-glass} 
&800 Hz& 0.75&1.69&8.416&12.62\\
&1250 Hz& 3.42&3.2&12.17&10.81\\ 
&2000 Hz& 14.58&9.01&22.92&9.38\\ 
\hline
\multirow{3}{*}{VCTK}
&  800 Hz&0.19 &0.53&1.25&1.44\\
&1250 Hz&1.6 &2.09&5.13&3.42\\ 
&2000 Hz& 3.24&4.15&5.64&3.83\\ 
\hline
\hline
 & &\multicolumn{2}{|c|}{ratio of recognizable}&\multicolumn{2}{|c}{longest chain}\\
\hline
dataset&frequency&Mean&SD&Mean&SD\\
\hline
\multirow{3}{*}{Pop-glass} 
&800 Hz& 0.02&0.06& 0.42&0.95\\
&1250 Hz& 0.25&0.26&2.17&1.82\\ 
&2000 Hz& 0.59&0.24&8.08&5.68\\ 
\hline
\multirow{3}{*}{VCTK}
&  800 Hz&0.07&0.2&0.19&0.53\\
&1250 Hz&0.26&0.34&1.4&1.89\\ 
&2000 Hz&0.4&0.41&3.0&4.0\\ 

\hline
\hline
 & &\multicolumn{2}{|c}{pairwise cosine similarity}&\multicolumn{2}{c}{}\\
\hline
dataset&frequency&Mean&SD&&\\
\hline
\multirow{3}{*}{Pop-glass} 
&800 Hz& 0.05&0.10& &\\
&1250 Hz& 0.02&0.04&&\\ 
&2000 Hz& 0.24&0.12&&\\ 
\hline
\multirow{3}{*}{VCTK}
&  800 Hz&0&0&&\\
&1250 Hz&0&0&&\\ 
&2000 Hz&0.05&0.13&&\\ 
\end{tabular}}
\label{table:metr}
\end{table}

\section{Conclusion} 
We investigated the privacy-preserving nature of low-frequency speech audio. While estimating voice activity is desirable for turn-taking dynamics in interactions, the ability to transcribe specific verbal content is a privacy risk. 
Our results indicate that 800 Hz and 2000 Hz are reasonable thresholds for maintaining VAD functionality whilst blocking intelligible content in clean and mingling-setting audio. 
Further, human intelligibility of bandwidth-extended low-frequency speech audio was slightly lower than an open-source ASR trained on web data, highlighting the challenges in transcribing such audio. 
While low-frequency recording shows promise in preserving privacy by obstructing intelligible speech, it is not a comprehensive solution. It remains an open question whether more advanced attacks might still extract sensitive information from low-frequency audio (e.g. model fine-tuning). 

\noindent\textbf{Acknowledgements}{
Thanks to Martha Larson for feedback on our final draft. This work was partially funded by the Erasmus+ funding program and the Netherlands Organization for Scientific Research, project number 639.022.606.}

\pagebreak

\bibliographystyle{IEEEtran}
\bibliography{mybib}

\end{document}